\documentclass{aastex}
\usepackage{emulateapj5}
\usepackage{graphicx}
\shorttitle{Thermal Conduction in MHD}
\shortauthors{Cho et al.}

\begin{document}
\title{Thermal Conduction in Magnetized Turbulent Gas}
\author{Jungyeon Cho and A. Lazarian}
\affil{Dept. of Astronomy, University of Wisconsin,
   Madison, WI53706; cho, lazarian@astro.wisc.edu}
\and
\author{Albert Honein, Bernard Knaepen, Stavros Kassinos, and Parviz Moin}
\affil{Center for Turbulence Research, Stanford University,
   Stanford, CA94305; honein, bknaepen, moin@stanford.edu}



\begin{abstract}
Using numerical methods, we systematically study in the framework of 
ideal MHD the effect of magnetic fields on heat
transfer within a turbulent gas.
We measure the rates of 
passive scalar diffusion within
magnetized fluids and make the comparisons a) between MHD and hydro
simulations, b) between different MHD runs with different values of
the external magnetic field (up to the energy equipartition value), 
c) between thermal conductivities parallel
and perpendicular to magnetic field.
We do {\it not} find apparent suppression of diffusion rates by the presence of
magnetic fields,
which implies that magnetic fields do not suppress heat diffusion by turbulent
motions.

\end{abstract}

\keywords{turbulence -- ISM: general -- galaxies: clusters: general -- MHD}

\section{Astrophysical Motivation}

It is well known that Astrophysical fluids are turbulent and that 
magnetic fields are dynamically important. 
One  characteristic of the medium that magnetic fields
and turbulence may substantially change 
is the heat transfer. 

There are many instances when heat transfer through thermal conductivity
is important. For instance, thermal conductivity is essential in
rarefied gases where radiative heat transfer is suppressed. This is exactly
the situation that is present in clusters of galaxies. It is widely
accepted that ubiquitous X-ray emission due to hot gas in clusters of
galaxies should cool significant amounts of the intracluster medium (ICM) and
this must result in cooling flows (Fabian 1994). However, observations
do not support the evidence for the cool gas (see Fabian et al. 2001)
which is suggestive of the existence of heating that replenishes the 
energy lost via X-ray emission. Heat transfer from the outer hot regions
can do the job, provided that the heat transfer is sufficiently efficient.

Gas in clusters of galaxies is magnetized and the conventional wisdom
suggests that the magnetic fields strongly suppress thermal conduction
perpendicular to their direction. Realistic magnetic fields are turbulent
and the issue of the thermal conduction in such a situation has been
long debated. A recent paper by Narayan \& Medvedev (2001) obtained
estimates for the thermal conductivity of turbulent magnetic fields,
but those estimates happen to be too low to explain the absence of 
cooling flows for many of the clusters of galaxies (Zakamska \& Narayan 2002).

Narayan \& Medvedev (2001) treat the turbulent magnetic
fields as static. In hydrodynamical turbulence
it is possible to neglect plasma turbulent motions only when the diffusion of
electrons which is the product of the electron thermal velocity $v_{elect}$
and the electron mean free path in plasma $l_{mfp}$, i.e. 
$v_{elect}l_{mfp}$, is greater than the turbulent velocity $v_{turb}$
times the turbulent injection scale $l_{inj}$, i.e. $v_{turb}l_{inj}$.
If such scaling estimates are applicable to heat transport in magnetized
plasma, the turbulent heat transport should be accounted for heat transfer
within clusters of galaxies. Indeed, data for
$v_{elect}l_{mfp}$ given in Zakamska \& Narayan (2002; Narayan \& Medvedev 2001)
provide the classical Spitzer (1962) diffusion 
coefficient
$\kappa_{Sp}\equiv v_{elect}l_{mfp} \sim 6.2 \times 10^{30}$ cm$^2$ sec$^{-1}$
for the inner region of $R\sim 100$kpc and 
$\kappa_{Sp}\equiv v_{elect}l_{mfp} \sim 3.6 \times 10^{29}$ cm$^2$ sec$^{-1}$
for the very inner region of $R\sim 10$kpc
(for Hydra A). 
If turbulence in the cluster of galaxies is of the order of
the velocity dispersion of galaxies, while the injection scale is
of the order of $20$~kpc, the diffusion coefficient is $\sim v_{turb}l_{inj}
\sim 3.1 \times 10^{30}$ cm$^2$ sec$^{-1}$, where we take
$v_{turb} \sim 500$ km/sec.

Earlier numerical studies by Cho, Lazarian \& Vishniac (2002) 
revealed a good correspondence between hydrodynamic motions and motions
of fluid perpendicular to the local direction of magnetic field. To what
extend heat transfer in a turbulent medium is affected by a magnetic field
is the subject of the present study. To solve this problem we shall
systematically study the passive scalar diffusion in a magnetized
turbulent medium, compare results of MHD and hydrodynamic calculations,
and investigate the heat transfer perpendicular and parallel to the mean
magnetic field for magnetic fields of different intensities.

This work has a broad astrophysical impact. Clusters of galaxies is
just one of the examples where non-radiative heat transfer is essential.
This process, however, is important for many regions within galactic
interstellar medium, e.g. for supernova remnants.

\section{Numerical methods}
We use a 3rd-order hybrid essentially non-oscillatory (ENO) upwind
shock-capturing scheme to solve the ideal MHD equations.
To reduce spurious oscillations near shocks, we
combine two ENO schemes.
When variables are sufficiently smooth, we use the 3rd-order
Weighted ENO scheme (Jiang \& Wu 1999)
without
characteristic mode decomposition.
When the opposite is true, we use the 3rd-order Convex ENO scheme 
(Liu \& Osher 1998).
We use a three-stage Runge-Kutta method for time integration.
We solve the ideal MHD equations in a periodic box:
\begin{eqnarray}
{\partial \rho    }/{\partial t} + \nabla \cdot (\rho {\bf v}) =0,  \\
{\partial {\bf v} }/{\partial t} + {\bf v}\cdot \nabla {\bf v}
   +  \rho^{-1}  \nabla(a^2\rho)
   - (\nabla \times {\bf B})\times {\bf B}/4\pi \rho ={\bf f},  \\
{\partial {\bf B}}/{\partial t} -
     \nabla \times ({\bf v} \times{\bf B}) =0,
\end{eqnarray}
with
    $ \nabla \cdot {\bf B}= 0$ and an isothermal equation of state.
Here $\bf{f}$ is a random large-scale driving force,
$\rho$ is density,
${\bf v}$ is the velocity,
and ${\bf B}$ is magnetic field.
The rms velocity $v_{turb}$ is maintained to be approximately unity, so that
 ${\bf v}$ can be viewed as the velocity
measured in units of the r.m.s. velocity
of the system and ${\bf B}/\sqrt{4 \pi \rho}$
as the Alfv\'{e}n velocity in the same units.
The time $t$ is roughly in units of the large eddy turnover time 
($\sim l_{inj}/v_{turb}$)
and the length in units of $l_{inj}$, the scale of the energy injection.
The magnetic field consists of a uniform background field and a
fluctuating field: ${\bf B}= {\bf B}_0 + {\bf b}$.

We use a passive scalar $\psi ({\bf x})$ to trace thermal particles.
We inject a passive scalar with a Gaussian profile:
\begin{equation}
   \psi({\bf x},t=t_0) \propto \exp^{-({\bf x}-{\bf x}_0)^2/\sigma_0^2}, 
   \label{eq_dist}
\end{equation}
where $\sigma_0$= $1/16$ of a side of the numerical box and ${\bf x}_0$ 
lies at the center of the computational box.
The value of $\sigma_0$ ensures that the scalar is injected
in the inertial range of turbulence.
The energy injection scale ($l_{inj}$) is $\sim 1/2.5$ 
of a side of the numerical box. 
The scalar field follows the continuity equation
\begin{equation}
{\partial \psi    }/{\partial t} + \nabla \cdot (\psi {\bf v}) =0.
\end{equation}

We are mainly concerned with time evolution of
$\sigma_i$ (i=x, y, and z):
\begin{equation}
  \sigma_i^2 = \frac{ \int (x_i-\bar{x}_i)^2 \psi({\bf x},t) d^3x }
                    { \int \psi({\bf x},t) d^3x  },
\end{equation}
where $\bar{x}_i=\left[ \int x \psi({\bf x},t) d^3x/\int \psi({\bf x},t) d^3x 
                 \right]$.
Common wisdom was that
the mean magnetic field suppresses diffusion
in the direction perpendicular to it.
If this is the case, we expect to see $\sigma_{\perp} < \sigma_{\|}$.
Otherwise, we will get $\sigma_{\perp} \sim \sigma_{\|}$.

We inject passive scalars after turbulence is fully developed.
Fig. \ref{fig_1}(a) shows when we inject the passive scalars.
For the hydrodynamic run with $M_s \mbox{ (sonic Mach number)}=0.3$ 
and $192^3$ grid points 
(thick solid line),
we inject passive scalars 5 times. The injection times are marked
by arrows.
We also mark the injection times by arrows 
for the MHD run with $V_A (=B_0/\sqrt{4\pi\rho})=1$, 
$M_s=0.3$, and $192^3$ grid points
(thin solid line for $<V^2>$ and dashed line for $<b^2>$).

\section{Theoretical Considerations}
Consider two massless particles in the inertial range.
Let the separation be $l$.
The separation follows
\begin{equation}
  \frac{ d l^2 }{ dt } \sim \frac{ (l+v_l dt)^2-(l-v_l dt)^2 }{ dt }
   \sim l v_l,
\label{eq_dl2dt}
\end{equation}
where we ignore constants of order unity.
Using $\epsilon \sim v_l^3 /l$, we get
\begin{equation}
        \frac{dl^2}{dt} \sim l (\epsilon l)^{1/3},
\end{equation}
where $\epsilon$ is the energy injection rate.
This leads to
\begin{equation}
  l^{2/3}-l_0^{2/3}  = (C_R)^{1/3} \epsilon^{1/3} (t-t_0),   \label{L23T}
\end{equation}
where $C_R$ is Richardson constant.
When $l \gg l_0$, we can write
\begin{equation}
  l^{2}  =  C_R \epsilon (t-t_0)^3,
\end{equation}
which was first discovered by Richardson (1926).

Recent direct numerical simulations suggest that $C_R \sim 1$.
Boffetta \& Sokolov (2002) obtained $C_R\sim 0.55$.
Ishihara \& Kaneda (2001) obtained $C_R\sim 0.7$.

When we inject a passive scalar field as in equation (\ref{eq_dist}), 
we may write
\begin{eqnarray}
  \sigma^{2/3}-\sigma_0^{2/3} 
              &=& (C_{1})^{1/3} \epsilon^{1/3} (t-t_0), \nonumber \\
              &=&  C_{2} v_{turb} (t-t_0),
        \label{S23T}
\end{eqnarray}
where $\sigma=( \sigma_x^2 + \sigma_y^2+ \sigma_z^2)^{1/2}$ and
the dimensionless constant $C_{1}$ is not necessarily 
the same as $C_{R}$. The constant
$C_2 \propto (C_1/l_{inj})^{1/3}$ has dimension.
In this paper, we do not attempt to obtain $C_1$ or $C_R$.
Instead, we investigate how $C_2$ behaves when we vary $B_0$.

Usually it was considered that
MHD turbulence is different from its hydrodynamic counterpart.
However, recently Cho, Lazarian, \& Vishniac (2002) showed that
motions perpendicular to the local mean fields are hydrodynamic to high order.
This means that many turbulent processes are as efficient as hydrodynamics ones.
For example, Cho et al. (2002) numerically showed that cascade timescale 
in MHD turbulence
follows hydrodynamic scaling relations (see also Maron \& Goldreich 2001).
The similarity between magnetized and unmagnetized turbulent flows
motivates us to speculate that turbulent mixing is
also efficient in MHD turbulence.
This is why we may use equation (\ref{S23T}), which is derived from hydrodynamic
turbulence.
It is worth noting that these facts are consistent with a recent model of
fast magnetic reconnection in turbulent medium (Lazarian \& Vishniac 1999).

\section{Results}

\begin{figure*}
\includegraphics[width=0.99\textwidth]{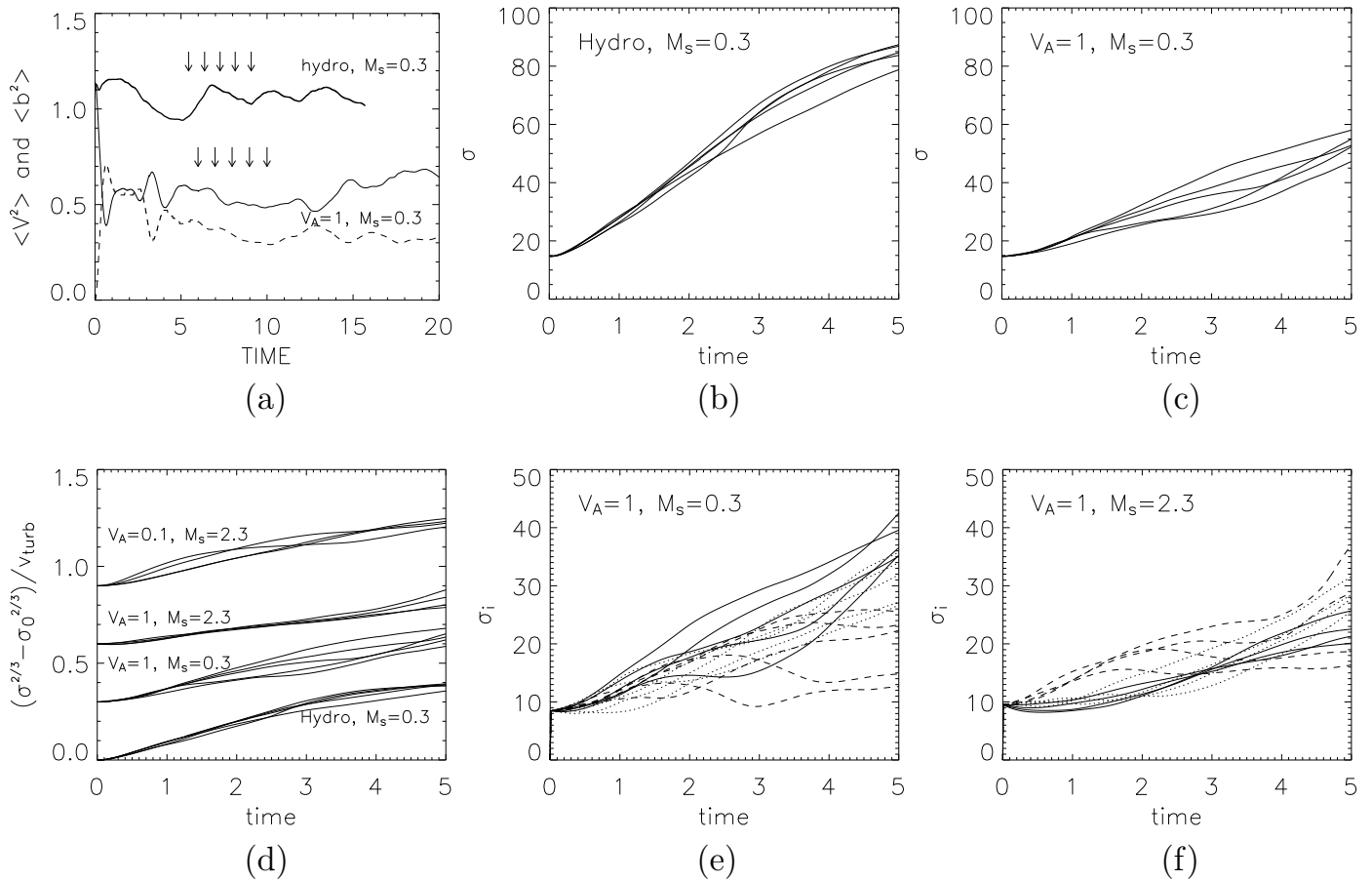}
\caption{ 
           Time evolution of energy density and $\sigma$.
  (a) Passive scalars are injected after turbulence is fully developed.
      Injection times are marked by arrows.
      Thick solid line: $<V^2>$ in the hydrodynamic run.
      Thin solid line: $<V^2>$ in the MHD run.
      Dashed line: $<b^2>$ in the MHD run.
  (b) Hydrodynamic run. $192^3$ grid points. $M_s$ (Mach no.) $\sim$ 0.3. 
  (c) MHD run. $192^3$ grid points. $M_s \sim$ 0.3.  $V_A=1$.
  (d) $(\sigma^{2/3}-\sigma_0^{2/3})/v_{turb}$ vs. time. 
        $\sigma$ is normalized by the box-size.
        Y-values are shifted by 0.3 units for convenience.
  (e) $\sigma_i$ (i=x, y, and z) vs. time.
      MHD run with $192^3$ grid points, $M_s\sim$ 0.3, and  $V_A=1$.
  (f) MHD run with $216^3$ grid points, $M_s\sim$ 2.3, and  $V_A=1$.
        Solid lines = x = parallel to ${\bf B}_0$;
        dashed lines = y; dotted lines = z.
    }
\label{fig_1}
\end{figure*}

In Figure \ref{fig_1}(b) and (c),
we compare the time evolution of $\sigma$ in hydrodynamic case and in
MHD case.
In the MHD case, the Alfven velocity of the mean field ($V_A=1$)
is slightly larger than the rms fluid velocity ($v_{turb}\sim 0.7$).
This is so-called subAlfvenic regime.
Since $V_A \sim v_{turb}$, the turbulence is strong.
The results show that turbulent diffusion is faster in hydrodynamic case.
However, Figure \ref{fig_1}(d) implies that this is due to
reduction in velocity.
Note that $v_{turb} \sim 1$ in the hydrodynamic case and $v_{turb} \sim 0.7$
in the MHD case
(see Figure \ref{fig_1}(a)).

Figure \ref{fig_1}(d) shows that there are good relations 
between $\sigma^{2/3}$ and $(t-t_0)$.
The slopes correspond to the constant $C_2$ in equation (\ref{S23T}).
The slopes are not very sensitive to $V_A$ or $M_s$.

Figure \ref{fig_1}(e) and (f)  shows that
diffusion rate does not strongly depend on the direction of the mean field.

The validity of equation (\ref{eq_dl2dt}) enables us to write
\begin{equation}
 \kappa_{dynamic} = C_{dyn} l_{inj} v_{turb},
\end{equation}
where $C_{dyn}$ is a constant of order unity.
This is the effective diffusion by turbulent motions suitable for scales
larger than $l_{inj}$.
The value of $C_{dyn}$ remains almost constant for $B_0$'s of up to
$B_0\sim \delta B \sim v_{turb}$.
The exact value of $C_{dyn}$ is uncertain.
In hydrodynamic cases, $C_{dyn}$ is of order of $\sim 0.3$ (see Lesieur 1990
chapter VIII and references therein).

\section{Astrophysical Implications}
We have shown that turbulence motions provide
 efficient mixing in MHD turbulence.
In this section, we show that
this process is as efficient as that proposed by 
Narayan \& Medvedev (2001) for some clusters.

\begin{figure*}
\begin{tabbing}
\includegraphics[width=0.25\textwidth]{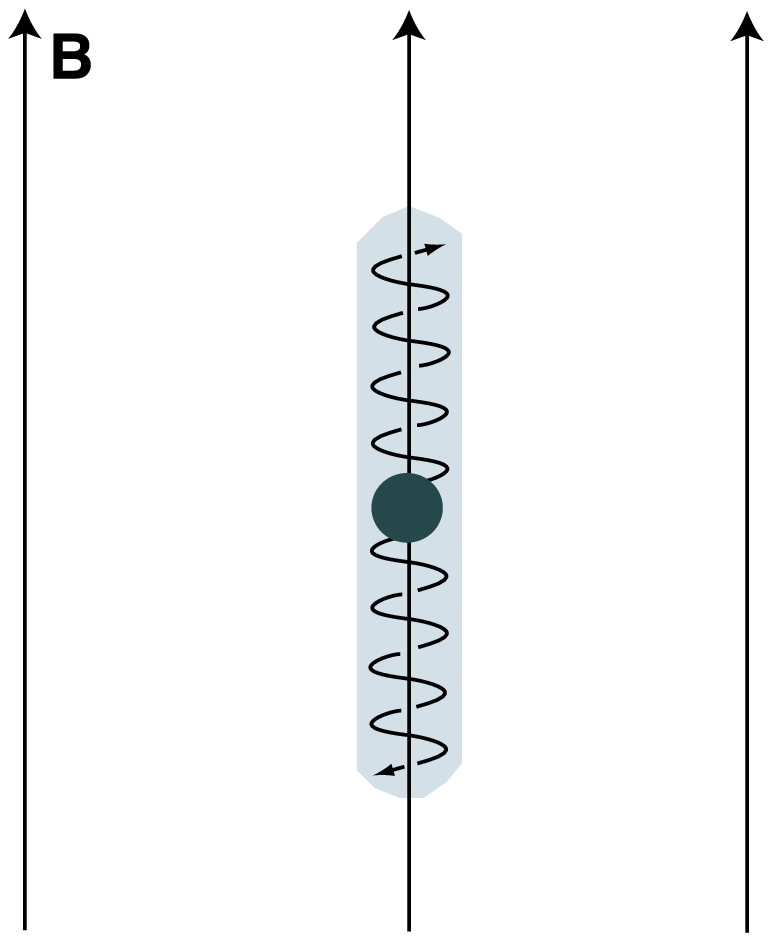}
\=
~~~~~~~\includegraphics[width=0.25\textwidth]{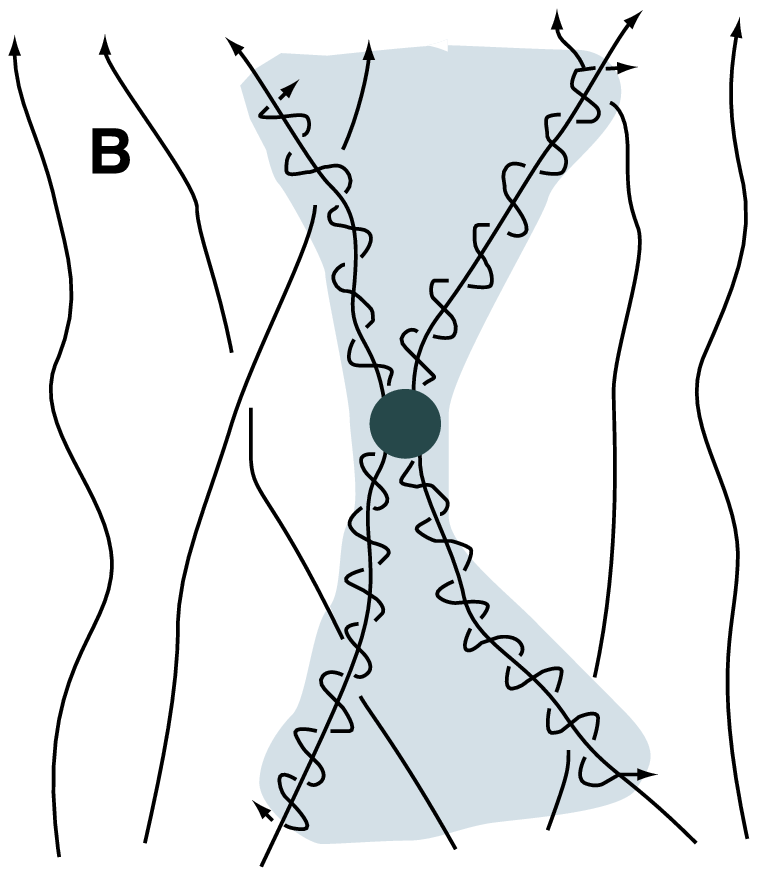}
\=
~~~~~~~ \includegraphics[width=0.30\textwidth]{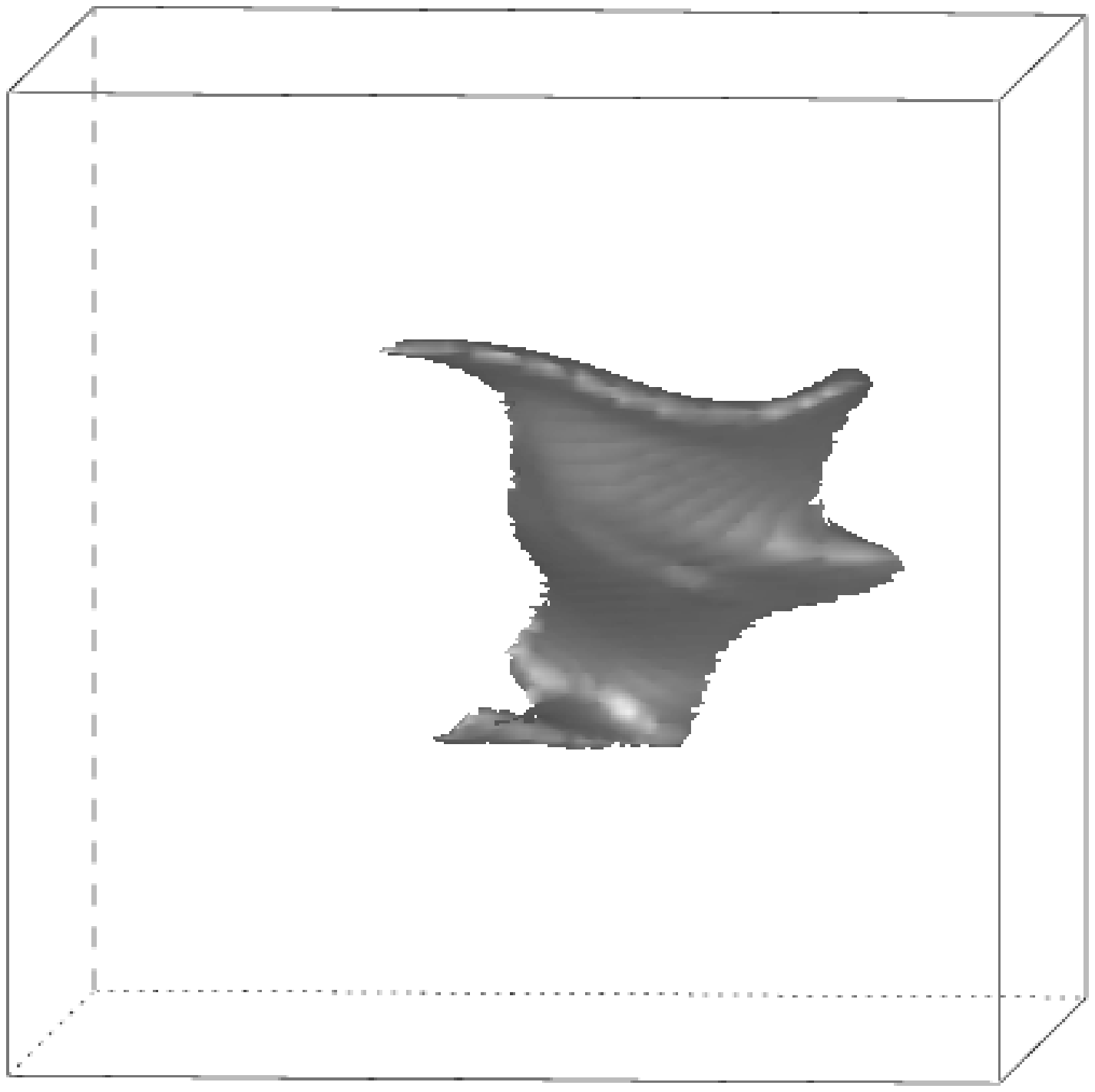} 
\\
~~~~~~~~~~~(a)  \> ~~~~~~~~~~~(b) \> ~~~~~~~~~~~(c) \\
\end{tabbing}
\caption{ Models of thermal diffusion. 
          (a) Classical picture. $\kappa_{\perp} \ll \kappa_{Sp}$.
          (b) Narayan \& Medvedev (2001). Wandering of field lines
              provides efficient diffusion ($\kappa_{\perp} \sim \kappa_{Sp}/5$)
              in the direction perpendicular
              to ${\bf B}_0$. But, the model assumes ${\bf B}_0$
              of $\sim$ equipartition value.
          (c) Turbulent diffusion model.
              Thermal electrons are mixed by turbulent motions, which
              leads to turbulent diffusion coefficient of 
              $\kappa_{dynamic}\sim v_{turb}l_{inj}$.
              In many astrophysical situations, this 
              coefficient is comparable with the
              Spitzer value. The figure is the
              snapshot of the passive scalar field at t$\sim$3 from
              the MHD run described in Figure 1(c);
              $192^3$ grid points, $M_s\sim$ 0.3, and  $V_A=1$.
              In the case shown here, the mean field is strong and 
              parallel to to the dashed line.
           In general, mean magnetic fields, weak or moderately strong, do not
              strongly suppress turbulent motions/diffusion. 
            }
\label{fig_3}
\end{figure*}

We summarize models of thermal diffusion in Fig.~\ref{fig_3}.
In the classical picture, thermal diffusion is highly suppressed in the
direction
perpendicular to ${\bf B}_0$.
Transport of heat along wondering
magnetic field lines (Narayan \& Medvedev 2001) 
partially alleviates the problem. 
But the applicability of Narayan \& Medvedev's model is a bit restricted - 
their model requires strong 
(i.e. $V_A \equiv B_0/\sqrt{4\pi \rho} \sim v_{turb}$) 
mean magnetic field.
In the Galaxy, there are strong mean magnetic fields. 
But, in the ICM, this is unlikely.
When the mean field is weak, the scales smaller than the characteristic
magnetic field scale ($\equiv l_{B}$) 
may follow the Goldreich \& Sridhar model (1995).
However, this requires further studies.
Our turbulent mixing model gives the same $\kappa_{dynamic}$
regardless of magnetic field geometry.

{\it ICM} ---
As we mentioned earlier, $\kappa_{Sp} \sim 1.3 \times 10^{30} (kT/keV)^{5/2}
(n/10^{-3}cm^{-3})^{-1} cm^2 sec^{-1}$ and 
$\kappa_{dynamic} \sim 3.1 \times 10^{30}
(v_{turb}/500km/s)(l_{inj}/20kpc) cm^2 sec^{-1}$.
The ratio of the two is of order unity for the ICM:
\begin{equation}
   \mu_{ICM} \equiv      \kappa_{dynamic} /\kappa_{Sp} \sim O(1).
\end{equation}
To be specific, for Hydra A, $\mu \sim 0.5$ 
for the inner region ($R\sim 100$kpc)
and $\mu \sim 8.6$ for the very inner region ($R\sim 10$kpc).
For 3C 295, $\mu \sim 0.34$ for the inner region and 
$\mu \sim 24$ for the very inner region.

{\it Local Bubble and SNRs} ---
The Local Bubble is a hot ($T\sim 10^6K; kT\sim$ 100 eV), 
tenuous ($n\sim 0.008/cm^3$) cavity
immersed in the interstellar medium (Berghofer et al 1998; Smith \& Cox 2001).
Turbulence parameters are uncertain.
We take typical interstellar medium values: $l_{inj}\sim$ 10 pc and
$v_{turb}\sim$ 5 km/sec.
For these parameters, the ratio of $\kappa_{dynamic}$ to $\kappa_{Sp}$ is
\begin{equation}
    \mu_{in} =       \kappa_{dynamic} /\kappa_{Sp} \sim  0.05,
\end{equation}
for the inside of the Local Bubble.
For the mixing layers, it is
\begin{equation}
    \mu_{mix} =       \kappa_{dynamic} /\kappa_{Sp} \sim  100,
\end{equation}
where we take $\bar{T}\sim \sqrt{T_cT_h}\sim 10^5 K$, 
$\bar{n}\sim \sqrt{n_cn_h}\sim 0.1/cm^3$ (Begelman \& Fabian 1990),
$T_c\sim 10^4 K$, $n_c \sim 1/cm^3$, 
$T_h\sim 10^6 K$, and $n_h \sim 0.008/cm^3$.
We expect similar results for supernova remnants 
since parameters are similar.

\section{Conclusion}
We have shown that magnetic fields (either random or mean  magnetic field
of up to equipartition value) do not suppress
turbulent diffusion processes,
which implies that turbulent diffusion coefficient has the form
$\kappa_{dyn} \sim l_{inj} v_{turb}$ in MHD turbulence, as well as
in hydrodynamic cases.
This result has two important astrophysical implications.
First, in the ICM, this turbulent diffusion coefficient is
of the same order of the classical Spitzer value.
Second, in the face of hot and cold media in the ISM (e.g. the boundary between
the Local Bubble and surrounding warm media),
this turbulent diffusion coefficient is much larger than the classical
Spitzer value.

\section*{Acknowledgments}
A.L. and J.C. acknowledge the support from the Center for Turbulence Research
(CTR)
at Stanford University.
This work 
utilized computing facilities at the CTR.

\clearpage

\end{document}